\documentclass[aps,prd,twocolumn,showpacs,superscriptaddress,letterpaper,asmmath,amssymb]{revtex4}

\usepackage{graphicx}
\usepackage{dcolumn}
\usepackage{bm}
\usepackage{feynmf}

\RequirePackage{xspace}
\usepackage{relsize}

\def\Kbar  {\kern 0.2em\overline{\kern -0.2em K}{}\xspace}

\def\Bbar    {\kern 0.18em\overline{\kern -0.18em B}{}\xspace}

\def\Qbar    {\kern 0.08em\overline{\kern -0.08em Q}{}\xspace}

\newcommand{\mev}{\ensuremath{\mathrm{\,Me\kern -0.1em V}}\xspace}
\newcommand{\mevc}{\ensuremath{{\mathrm{\,Me\kern -0.1em V\!/}c}}\xspace}
\newcommand{\mevcc}{\ensuremath{{\mathrm{\,Me\kern -0.1em V\!/}c^2}}\xspace}
\newcommand{\gev}{\ensuremath{\mathrm{\,Ge\kern -0.1em V}}\xspace}
\newcommand{\gevc}{\ensuremath{{\mathrm{\,Ge\kern -0.1em V\!/}c}}\xspace}
\newcommand{\gevcnospace}{\ensuremath{{\mathrm{\,Ge\kern -0.1em V\!/}c}}}
\newcommand{\gevcc}{\ensuremath{{\mathrm{\,Ge\kern -0.1em V\!/}c^2}}\xspace}
\newcommand{\bea}{\begin{eqnarray}}
\newcommand{\eea}{\end{eqnarray}}
\newcommand{\del}{\partial}

\begin{document}
\setlength{\unitlength}{1mm}
\bibliographystyle{apsrev}

\title{Searching for $Z'$ bosons decaying to gluons}

\author{Johan~Alwall}
\affiliation{Fermi National Accelerator Laboratory, Batavia, IL 60615}
\affiliation{National Taiwan University, Dept. of Physics, Taipei 10617 , Taiwan}
\author{Mazin~Khader}
\affiliation{University of California, Irvine, Irvine, California 92697}
\author{Arvind Rajaraman}
\affiliation{University of California, Irvine, Irvine, California 92697}
\author{Daniel~Whiteson}
\affiliation{University of California, Irvine, Irvine, California
  92697}
\author{Michael~Yen}
\affiliation{University of California, Irvine, Irvine, California
  92697}

\begin{abstract}
The production and decay of a new heavy vector boson, a chromophilic
$Z'$ vector boson, is described. The chromophilic $Z'$ couples only
to two gluons, but its two-body decays are absent, leading to a
dominant decay mode of $Z'\rightarrow q\bar{q}g$.  The unusual 
nature of the interaction predicts a
cross-section which grows with $m_{Z'}$ for a fixed coupling and an accompanying gluon with a coupling that rises with its
energy. We study the
$t\bar{t}g$ decay mode, proposing distinct reconstruction techniques
for the observation of an excess and for the measurement of
$m_{Z'}$. We estimate the sensitivity of current experimental datasets.
\end{abstract}



\maketitle

\date{\today}

\section{Introduction}


Many models of physics beyond the Standard Model predict the existence of new
$U(1)$ gauge factors (e.g.\cite{Leike:1998wr,Rizzo:2006nw,Langacker:2008yv,Nath:2010zj}). For example, 
grand unified theories with SO(10) gauge group
naturally have an extra $U(1)$ gauge factor~\cite{Ross:1985ai}. Models with extra dimensions at the
TeV scale can have extra gauge factors on the hidden brane. String theoretic
models usually have extra branes wrapped around higher
dimensional cycles, as well as intersecting branes, which can produce new gauge factors
~\cite{Blumenhagen:2000vk,Cvetic:2001nr,Cvetic:2002wh,Uranga:2003pz,Cvetic:2004nk,Marchesano:2004yq,Cvetic:2004xx,Kumar:2005hf,Kumar:2007zza,Douglas:2006xy}.
In most of these cases, these new $U(1)$ gauge factors are
typically broken either by
the Green-Schwarz mechanism or by a charged scalar expectation value, so that the
corresponding gauge boson is massive.

If the new sector is completely secluded from the Standard Model, it does not
have phenomenological consequences. However, in many of these models, there are massive fields charged
under both the hidden and visible gauge groups. 
Once these fields are integrated out, they can induce couplings
between
the hidden and visible sectors, which are observable at colliders.
This has motivated a great deal of effort in searches for new
gauge fields, and in particular new $Z'$ gauge bosons.
If the $Z'$ boson couples to quarks and leptons, it can produce spectacular signals at colliders
as a dijet or dilepton resonance. Current colliders already place stringent constraints
on such new bosons which have coupling similar to the Standard Model
$Z$ boson~\cite{Alcaraz:2006mx,Jaffre:2009dg,Chatrchyan:2011wq}.

It is however, very plausible that these new gauge bosons have highly suppressed
direct couplings to quarks
and leptons. If the new gauge boson is from a hidden sector as in string-theoretic models
or in models where dark matter arises from a hidden sector, there
are typically no tree-level couplings between the Standard Model fermions and the
$Z'$ boson.  At the loop level, there can be quantum corrections that
mix the $Z'$ boson
with the $U(1)$ of hypercharge~\cite{Holdom:1985ag,delAguila:1995rb,Dienes:1996zr,Kumar:2006gm,Feldman:2006wb,Chang:2006fp,Chang:2007ki,Feldman:2007wj}. These are called kinetic mixing terms; they are
renormalizable and hence unsuppressed by a mass scale.  These couplings then
induce a coupling between the
$Z'$ boson and the Standard Model fermions.

However, if there are no fields charged under both
hypercharge and the new $U(1)'$, these kinetic mixing terms are
 absent.
In this case, the leading interactions between the hidden sector
and the Standard Model will come from bifundamental fields charged under
$U(1)'$ and either $SU(2)$ or $SU(3)$. Once these fields are
integrated out, there will be new couplings induced between
the hidden sector and the gauge bosons of the visible sector which are  of
the form $Z' G^2$, where $G$ is a field strength either of $SU(2)$ or $SU(3)$.

In this paper, we shall consider the case where the $Z'$ boson is coupled to the $SU(3)$ field
strength (the case where the coupling is only to the $SU(2)$ field strength was considered
in~\cite{su2}). We shall refer to these as chromophilic $Z'$ bosons. We shall discuss the current 
constraints on
such
models and possible searches for these models in current experimental collider datasets. We shall
be interested to see if this model can
be discovered at the Tevatron or the LHC.

Specifically, we will  consider a hidden sector consisting of a $U(1)$
theory broken by an abelian Higgs model. The physical
spectrum will then have one massive gauge boson
which we denote as $Z'$. This sector
is coupled to the Standard Model by
mediator fields $\psi$ charged under the $SU(3)$ of the Standard Model as well as the hidden sector
$U(1)$. The resulting operators will depend on whether the $Z'$ boson is a vector or a pseudovector.
Motivated by string-theoretic models, we consider the case where $Z'$ boson is a pseudovector.
For an on-shell $Z'$ boson, the only relevant operator is then~\cite{su2}
\bea
{\cal L}_{int}=
g \epsilon_{\mu\nu\rho\sigma} Z^\mu G^{\nu\rho}\del_\alpha G_{\alpha\sigma}
\label{eq:1}
\eea
\noindent
where $g$ has dimensions of mass$^{-2}$. We aim to find the sensitivity of current experiments as a function
of $Z'$ boson mass and the coupling $g$.

Despite the $Z'$-gluon-gluon vertex, the chromophilic $Z'$ boson has no two-body decays.
This is because the Landau-Yang theorem~\cite{LandauYang} prevents a massive gauge boson from 
decaying to
two massless gauge bosons. 
The only possible decays are three-body decays; the $Z'$ boson can decay to
two quarks and a gluon through an off-shell gluon
(we found by explicit calculation that the $Z'$ boson decay to three gluons also vanishes).
Furthermore, the $Z'$ boson is not produced directly in the process $gg\to Z'$ for the same
reason.

The leading production process is through the process $qg\to qZ'$ or
$gg\to gZ'$ through an off shell gluon, followed by the decay $Z'\to
q\bar{q}g$. This leads to a ($q\bar{q}qg$ or $q\bar{q}gg$) final
state; if the $q\bar{q}$ pair are light, it gives a four-jet final
state which is challenging to see over the large multi-jet background.
The usual constraints on $Z'$ models from dilepton and dijet final states
therefore do not apply to this model, which would appear instead in
events with four jets.

\begin{figure}[floatfix]
\begin{center}
\includegraphics[width=2.5in]{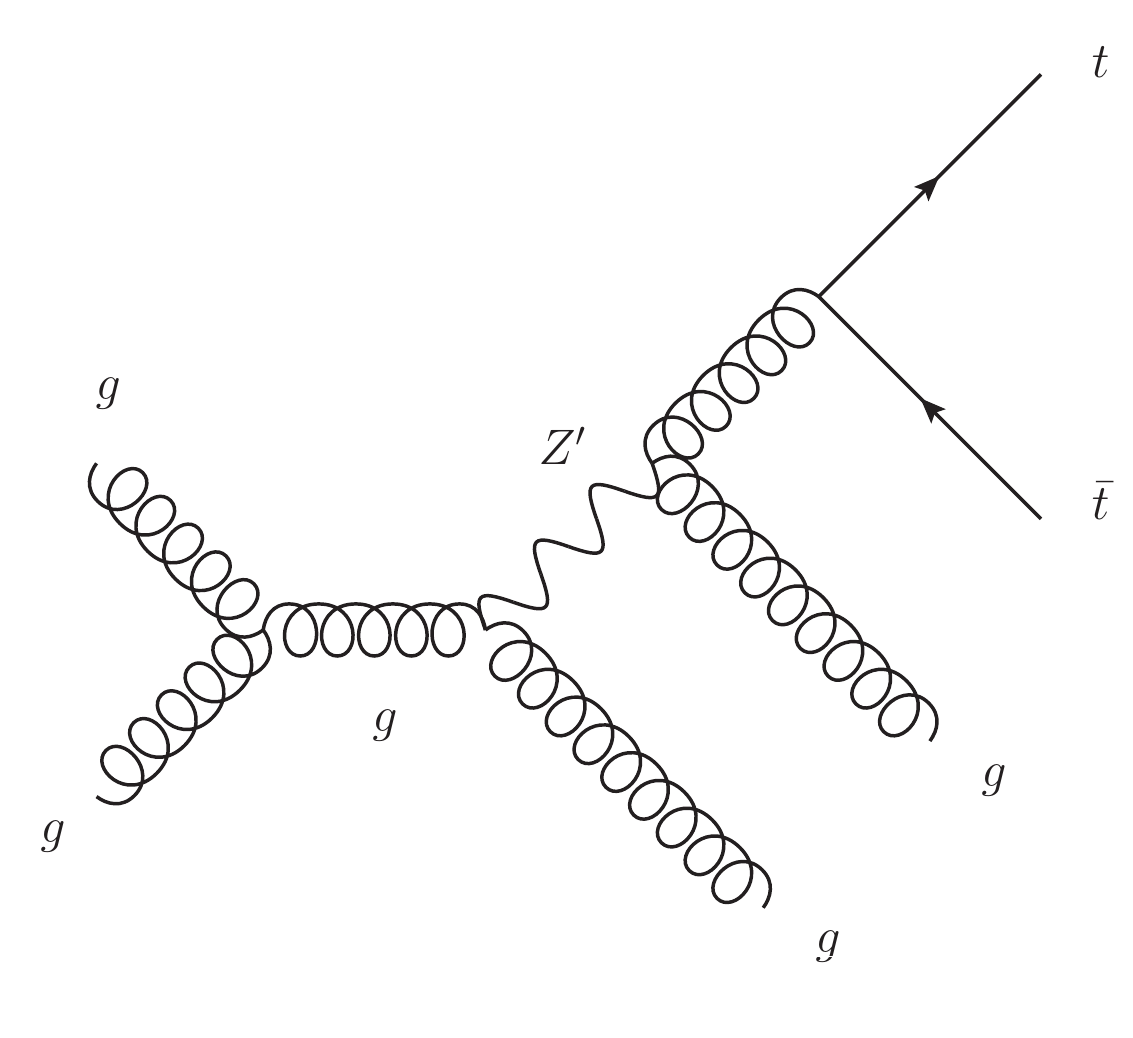}
\includegraphics[width=2.5in]{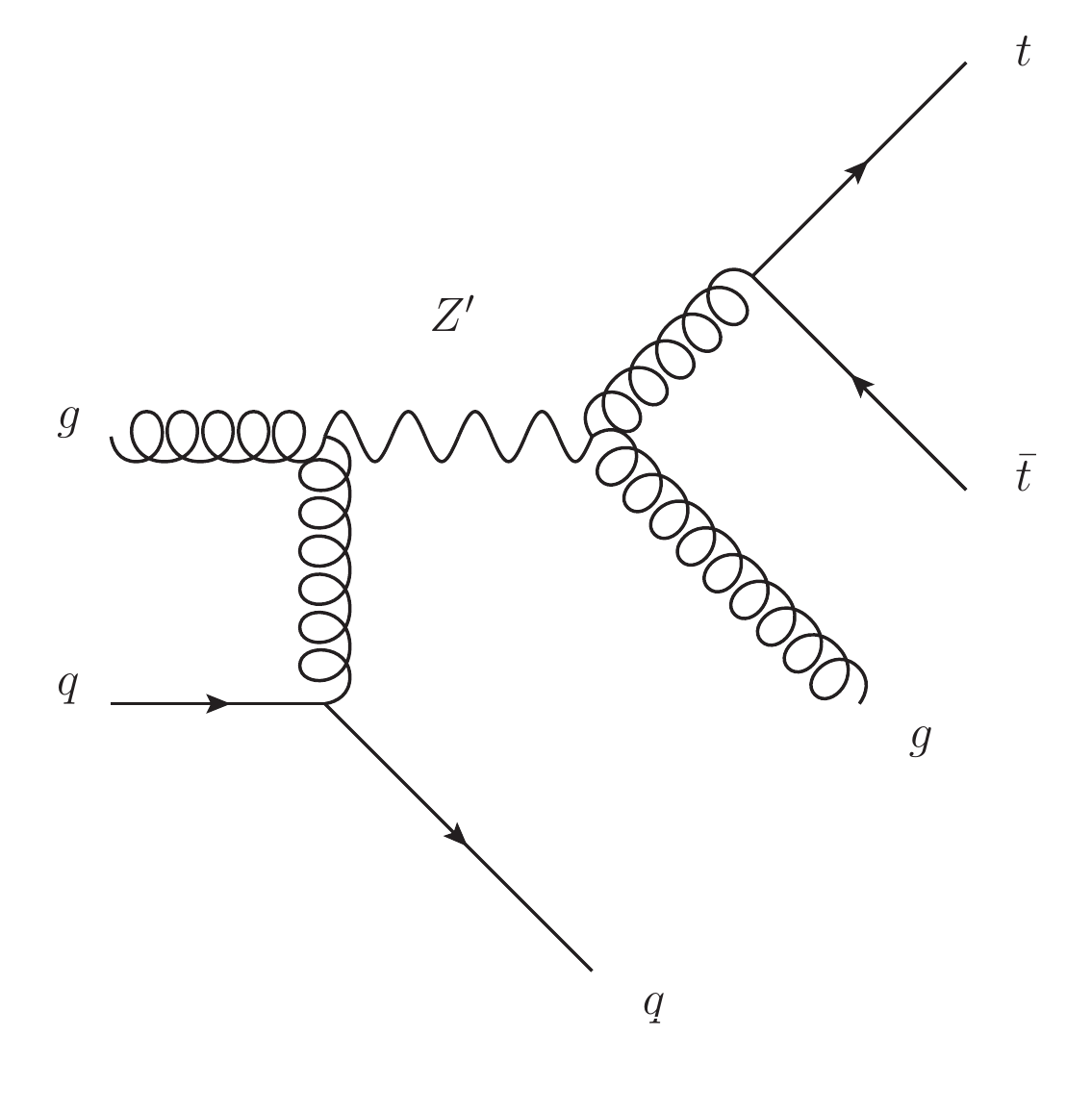}
\end{center}
\caption{ Diagram for $Z'g$ (top) or $Z'q$ (bottom) production followed by $Z'\rightarrow gg^*
  \rightarrow g t\bar{t}$ decay giving a $t\bar{t}gg$ (top) or
  $t\bar{t}gq$ (bottom) final state.}
\label{fig:diagram}
\end{figure}

To extract the signal from the large background, we will
look at signal events with heavy flavor.
In this paper, we focus on the decay  $Z'\rightarrow gt\bar{t}$ (see
Figure~\ref{fig:diagram}), which gives a final state of $t\bar{t}+2$ jets.
We will be aided by the fact that we can require the two heavy quarks
along with one of the other jets to reconstruct to the $Z'$ resonance
($t\bar{t}j$). This final state signature, $t\bar{t}+2$ jets, with a resonance in
$t\bar{t}j$ has not yet been experimentally explored.


\section{Selection and Backgrounds}

The event selection roughly follows the standard selection for
$t\bar{t}\rightarrow \ell$+jets analyses~\cite{ljets1,ljets2} except
that we require a fifth jet. Briefly, we
require:

\begin{itemize}
\item exactly one electron or muon, with $p_T>20$ GeV and $|\eta|<2.5$
\item at least five jets, each with $p_T>20$ GeV and $|\eta|<2.5$
\item at least 20 GeV of missing transverse momentum
\item at least one $b$-tagged jet
\end{itemize}

The dominant Standard Model background is $t\bar{t}$ production with additional jets from
initial- or final-state
radiation. At the Tevatron (LHC) , $W+$jets contributes 25\% (10\%). In
this study, we consider only the $t\bar{t}$ background.

Both the signal and background events are generated with
{\sc madgraph 5}~\cite{madgraph5}, while top-quark and $W$-boson decay,
showering and hadronization are performed by {\sc pythia 6.4}~\cite{pythia}.
The parametric detector simulation program {\sc pgs}~\cite{pgs} is  tuned for Tevatron or ATLAS as appropriate.

The expected background levels are calculated using the NLO cross-section~\cite{ttbarnlo}
for  $t\bar{t}+j$ production, acceptance calculated with simulated events, and a
luminosity of 8 fb$^{-1}$ (5 fb$^{-1}$) for the Tevatron (LHC). A 10\%
normalization uncertainty is used.  The acceptance for $Z'q$ and $Z'g$ production  is calculated using
simulated events. Table~\ref{tab:acc} shows the acceptances of signal
and background for both datasets.

\begin{table}[floatfix]
\caption{Acceptance of the event selection  for
  $Z'+j$ production and the dominant background, SM
  $t\bar{t}+2j$. Statistical uncertainty is approximately 1\%.}
\label{tab:acc}
\begin{tabular}{lcc}
\hline \hline
& \multicolumn{2}{c}{Acceptance} \\
& Tevatron & LHC  \\
& $p\bar{p}$, 1.96 TeV & $pp$, 7 TeV \\ \hline
SM $t\bar{t}+1j$ & 5\%  &  11\% \\
$Z'+j$ ({\tiny $400$ GeV}) & 7\% & 9\%\\
$Z'+j$ ({\tiny $500$ GeV}) & 9\% & 8\%\\
$Z'+j$ ({\tiny $600$ GeV}) & 11\%  & 9\%\\
$Z'+j$ ({\tiny $700$ GeV}) & 11\%  & 10\%\\
$Z'+j$ ({\tiny $800$ GeV}) & 12\%  & 10\%\\
$Z'+j$ ({\tiny $900$ GeV}) & 12\% & 11\%\\
$Z'+j$ ({\tiny $1000$ GeV}) & 12\%  & 11\%\\

\hline \hline
\end{tabular}
\end{table}

\section{Reconstruction and Sensitivity}

\begin{figure}[floatfix]
\begin{center}
\includegraphics[width=0.45\linewidth]{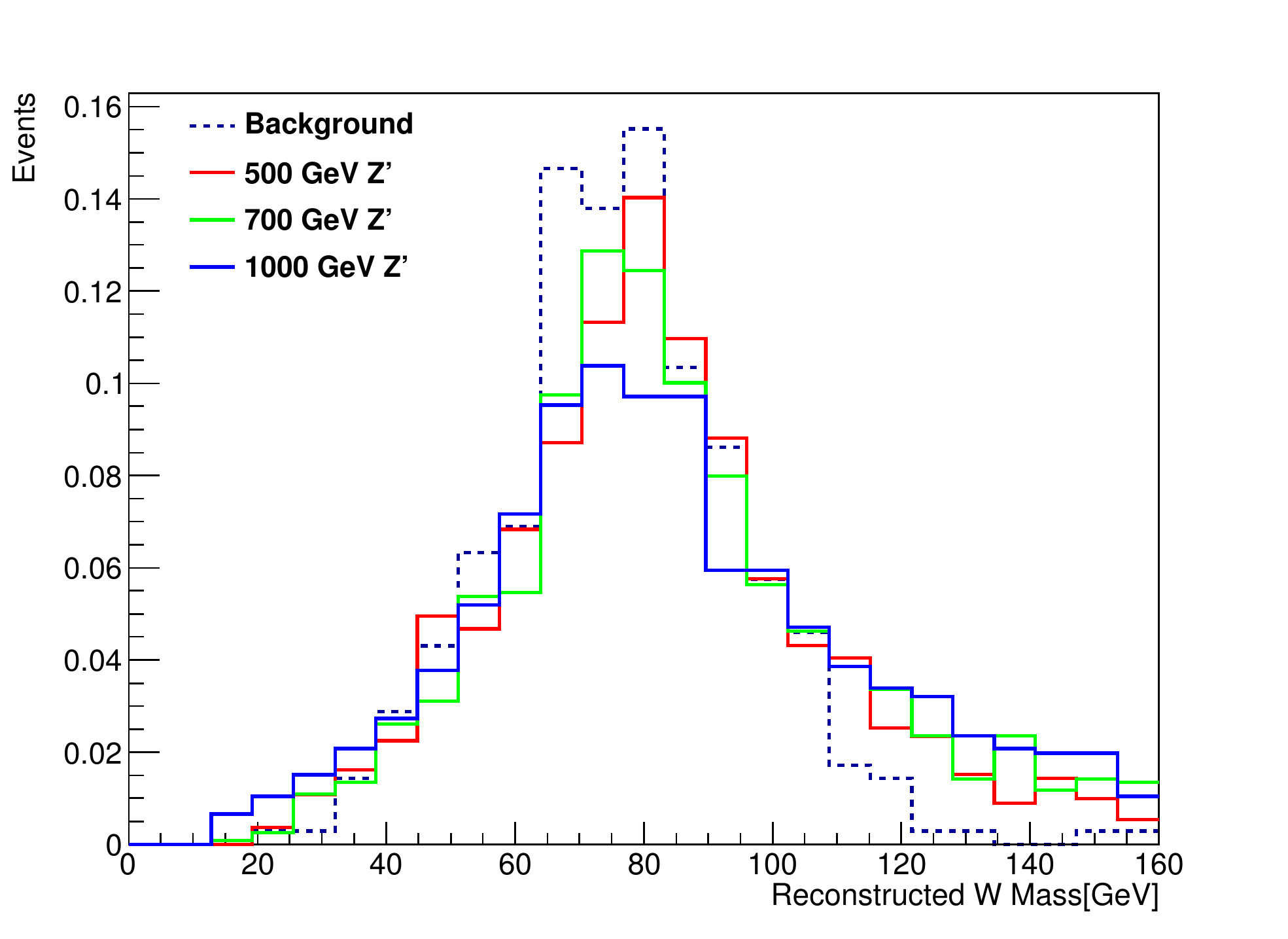}
\includegraphics[width=0.45\linewidth]{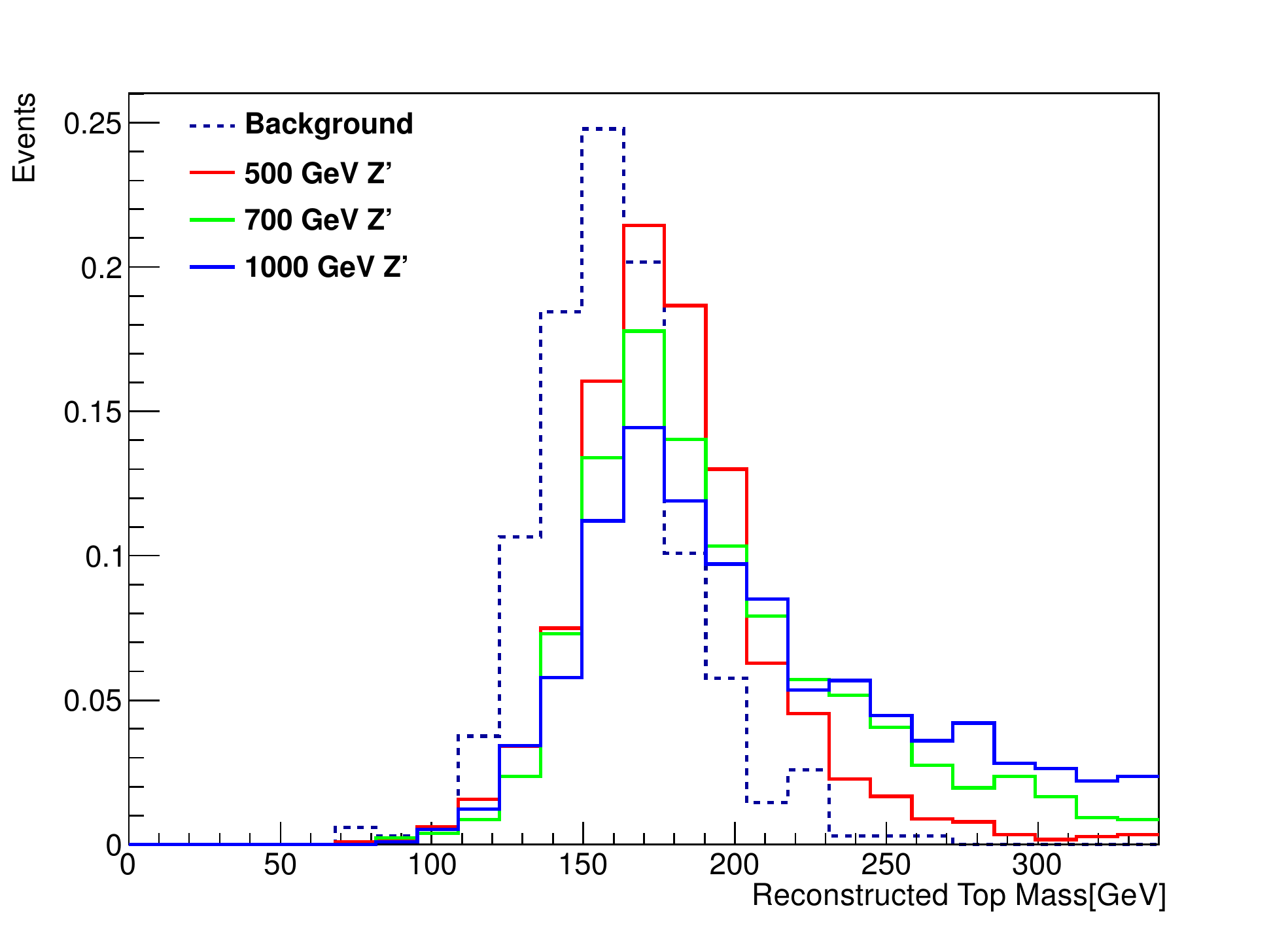}\\
\includegraphics[width=0.45\linewidth]{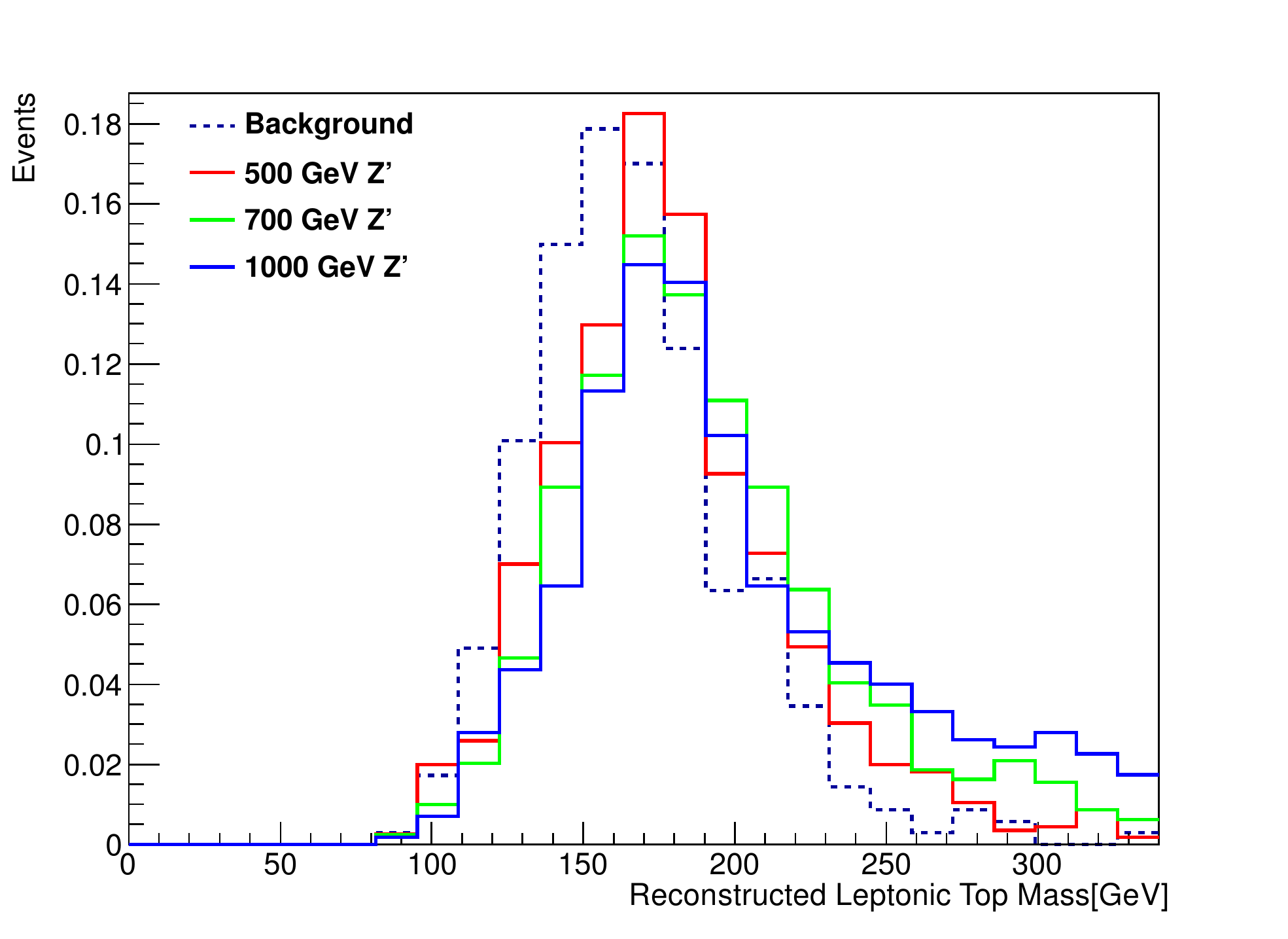}
\includegraphics[width=0.45\linewidth]{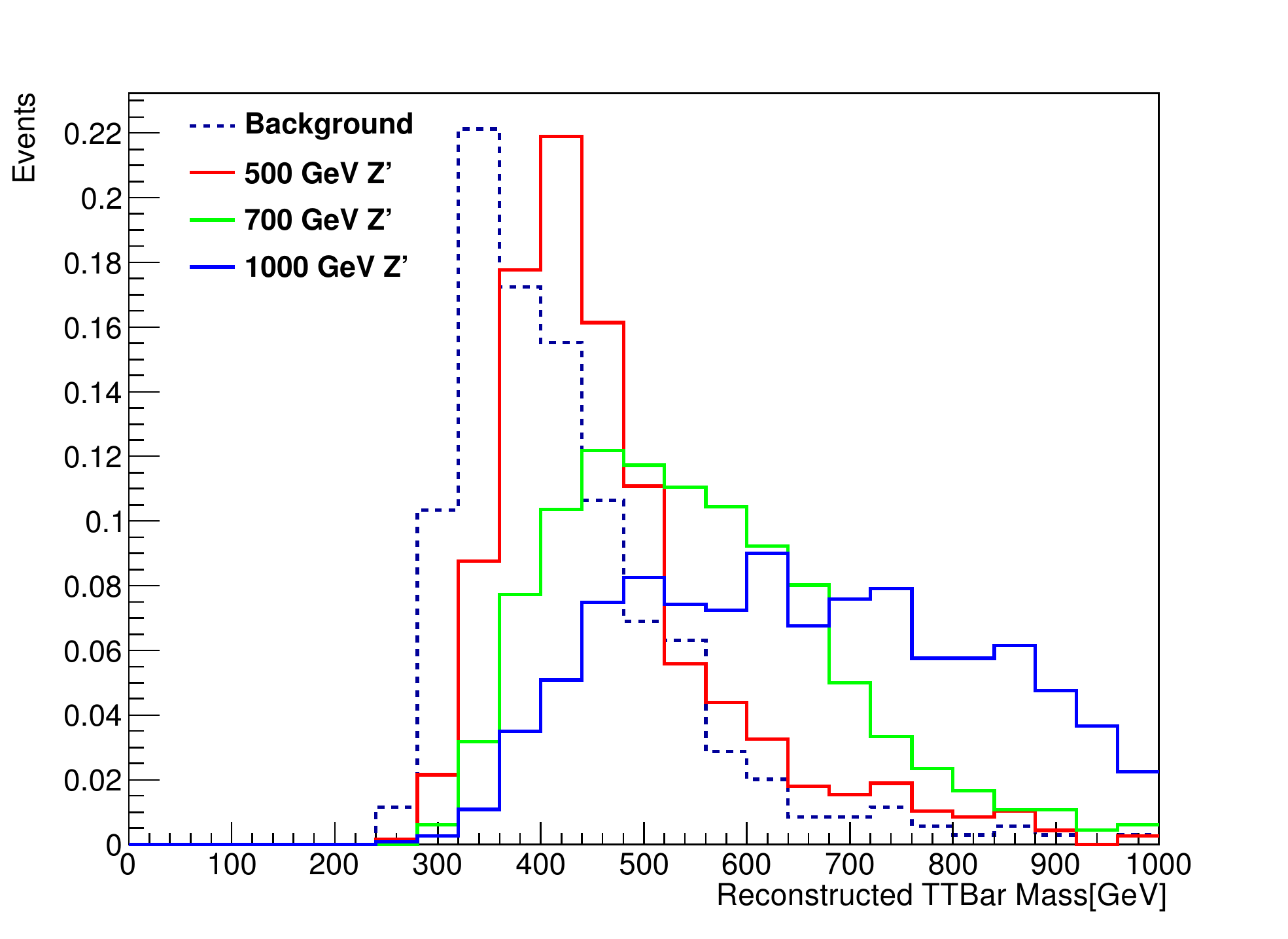}
\end{center}
\caption{ Distribution at the Tevatron of $m_{qq'}, m_{qq'b}, m_{\ell\nu b'}$, and
  $m_{t\bar{t}}$ for the dominant SM background of $t\bar{t}$+jets and
  for two choices of $Z'$ signal.}
\label{fig:tev}
\end{figure}

\begin{figure}[floatfix]
\begin{center}
\includegraphics[width=0.45\linewidth]{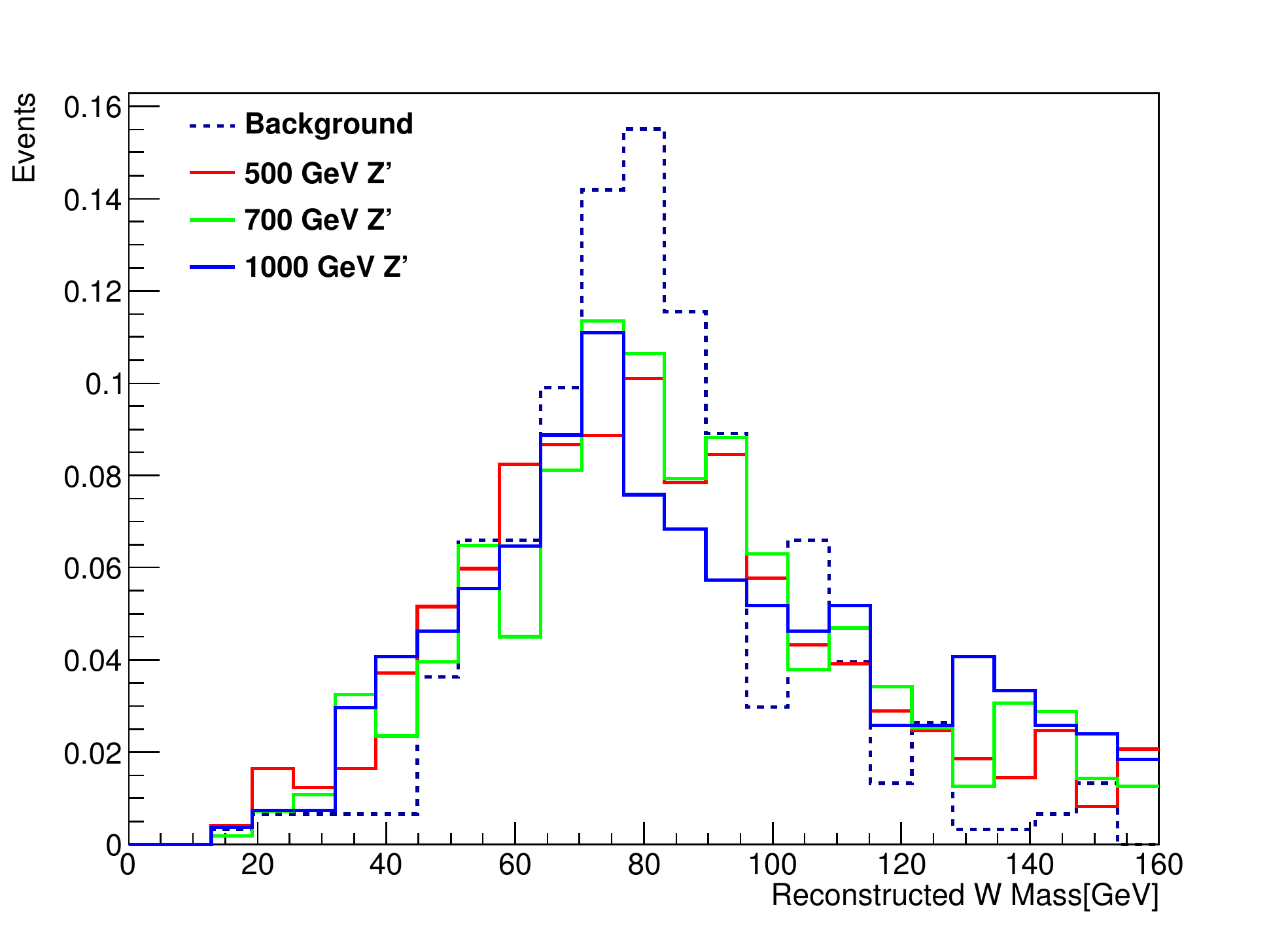}
\includegraphics[width=0.45\linewidth]{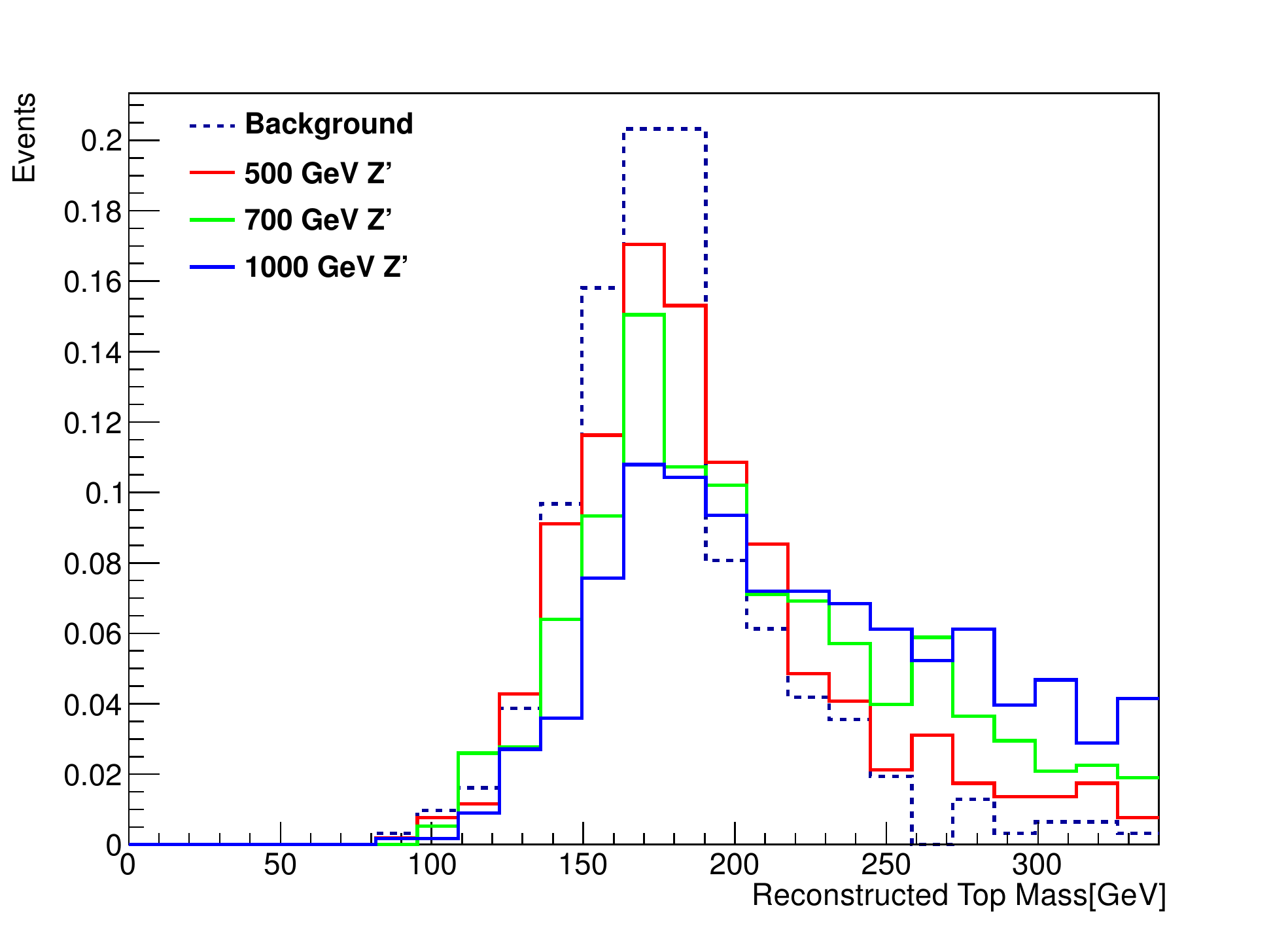}\\
\includegraphics[width=0.45\linewidth]{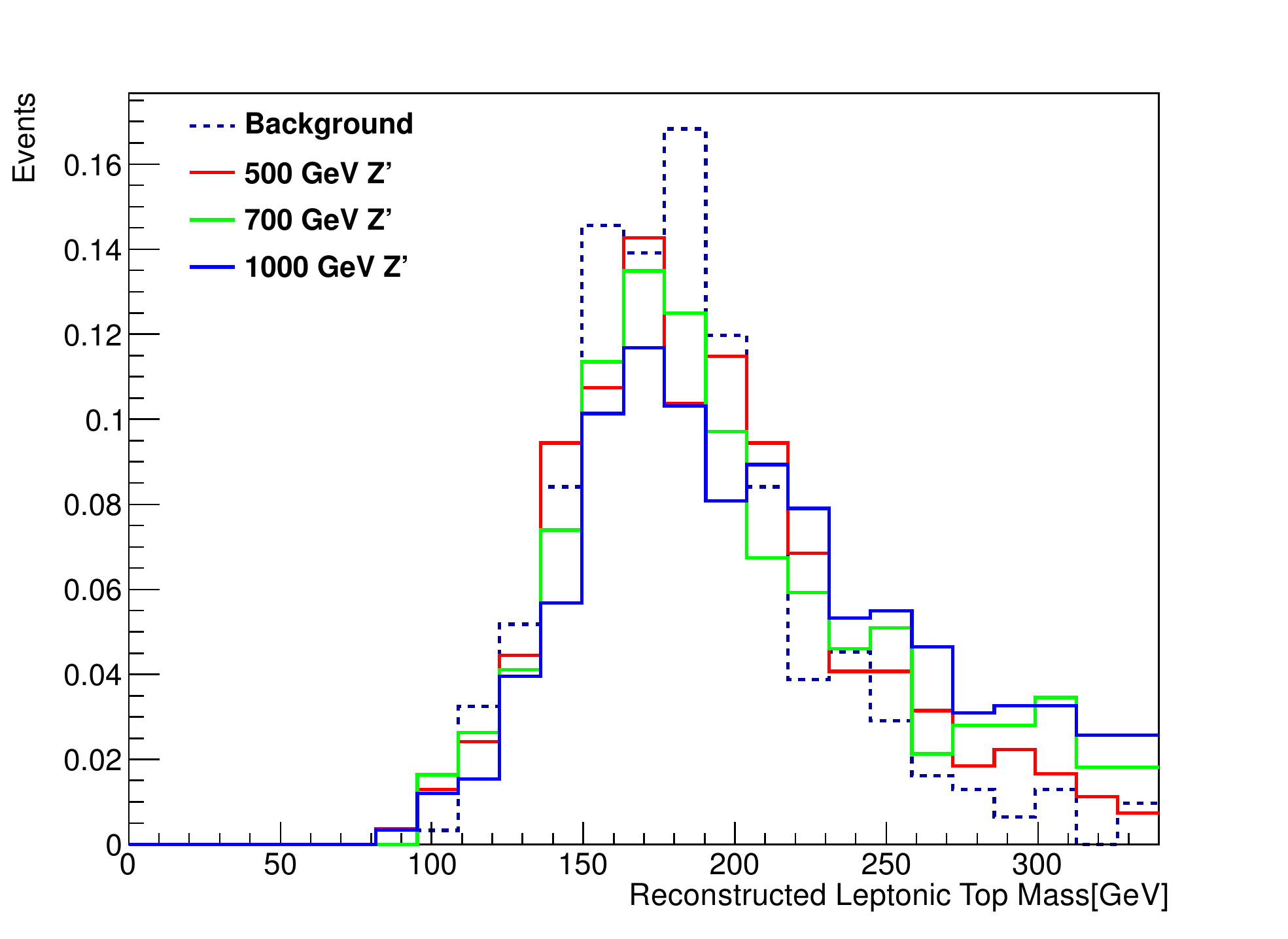}
\includegraphics[width=0.45\linewidth]{mttbar_tev.pdf}
\end{center}
\caption{ Distribution at the LHC of $m_{qq'}, m_{qq'b}, m_{\ell\nu b'}$, and
  $m_{t\bar{t}}$ for the dominant SM background of $t\bar{t}$+jets and
  for two choices of $Z'$ signal.}
\label{fig:lhc}
\end{figure}

Events are reconstructed according to the $t\bar{t}$ hypothesis. The neutrino
transverse momentum is taken from the missing transverse momentum; the
longitudinal component is set to the smallest value which gives
$(p_\ell+p_\nu)^2=m_W^2$. The jets from hadronic $W\rightarrow qq'$ decay and
the two $b$-quarks are identified by selecting the jets which minimize
the function:

\[ \chi^2 = \frac{ (m_{qq'} - m_W)^2}{\sigma_{qq'}^2} + \frac{ (m_{qq'b}
  - m_t)^2}{\sigma_{qq'b}^2} + \frac{ (m_{\ell\nu b'} -
  m_t)^2}{\sigma_{\ell\nu b'}} \]

\noindent
where the denominator $\sigma_{qq'},\sigma_{qq'b},\sigma_{\ell\nu b}$ values
which describe the resolution of each mass term are extracted from
simulated events.  All jets which satisfy the $p_T$ and $\eta$
requirements above are considered.   Distributions of reconstructed
$W$ boson and top quark
candidate masses are
shown in Figures~\ref{fig:tev} and~\ref{fig:lhc} and demonstrate that the
reconstruction accurately identifies the $W$-boson and top-quark decays.

\begin{figure}[floatfix]
\begin{center}
\includegraphics[width=0.45\linewidth]{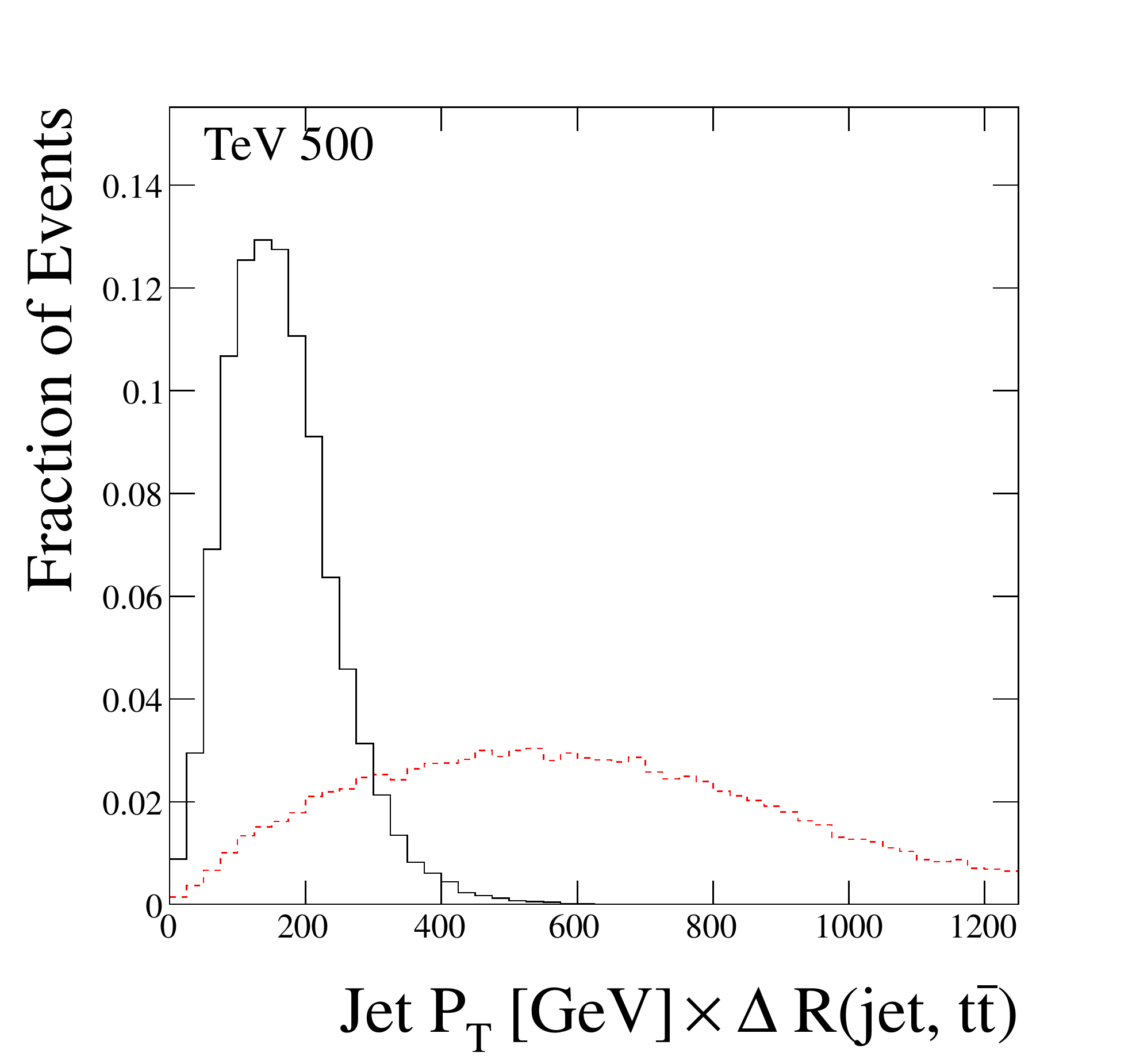}
\includegraphics[width=0.45\linewidth]{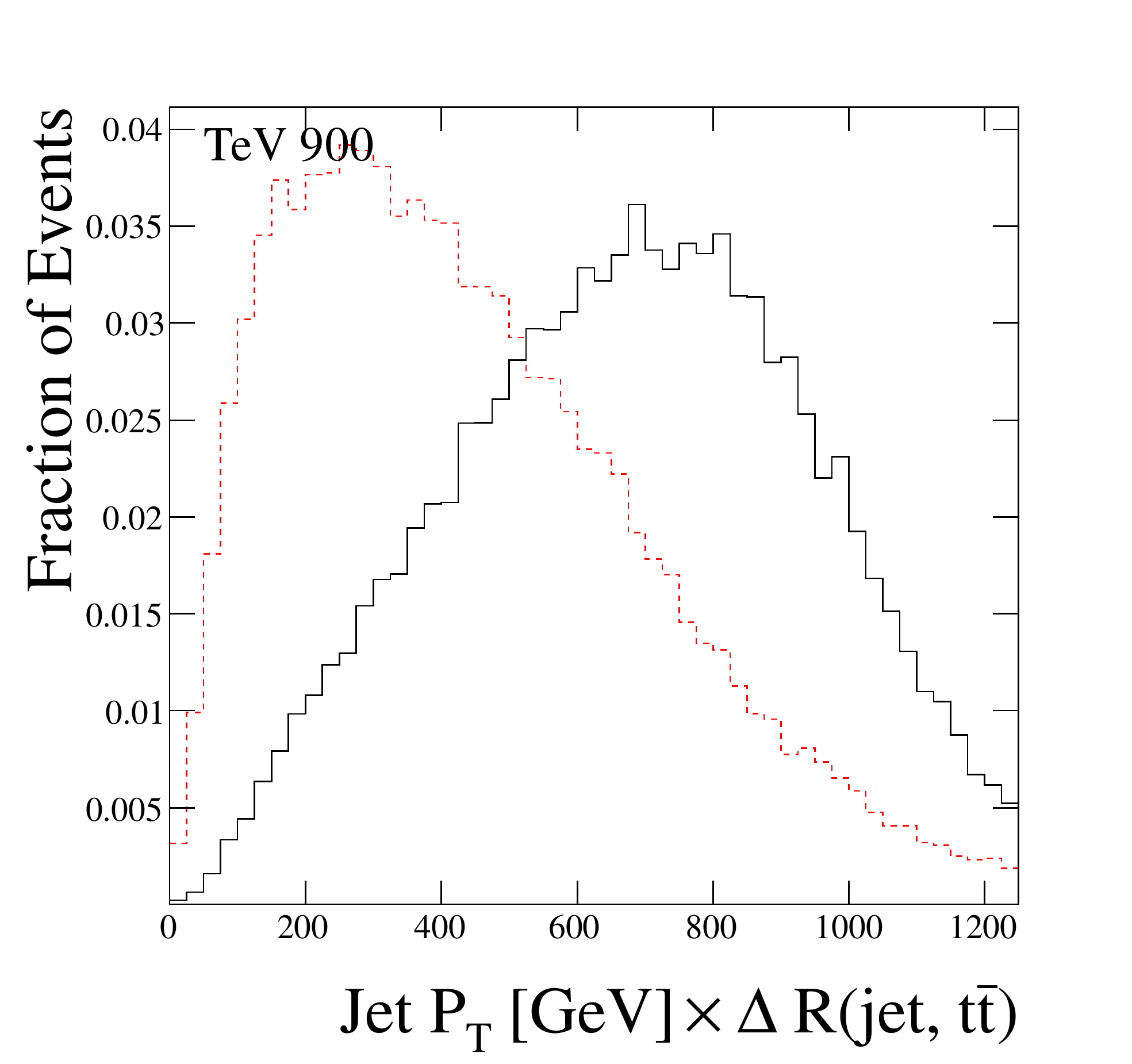}
\includegraphics[width=0.45\linewidth]{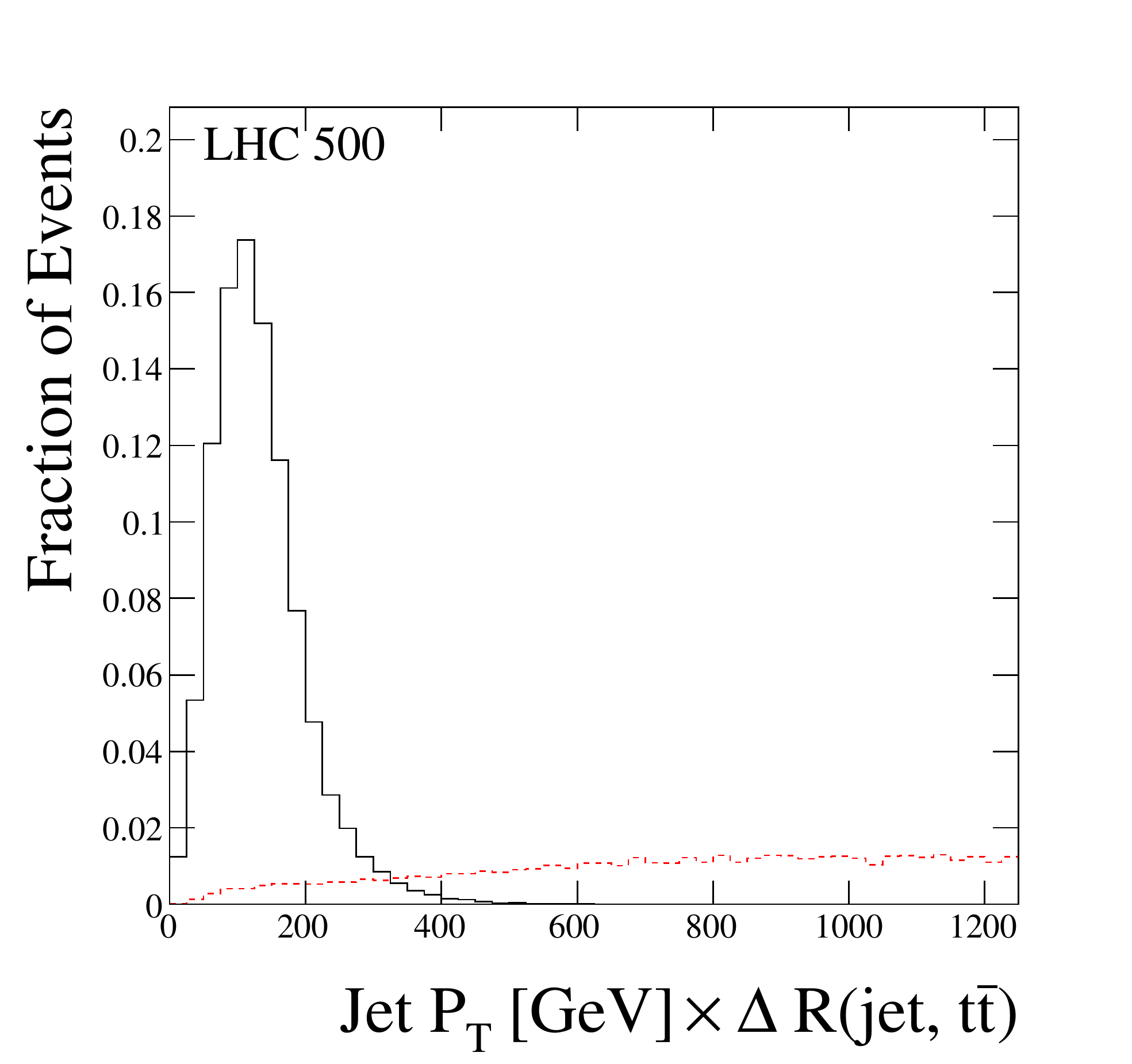}
\includegraphics[width=0.45\linewidth]{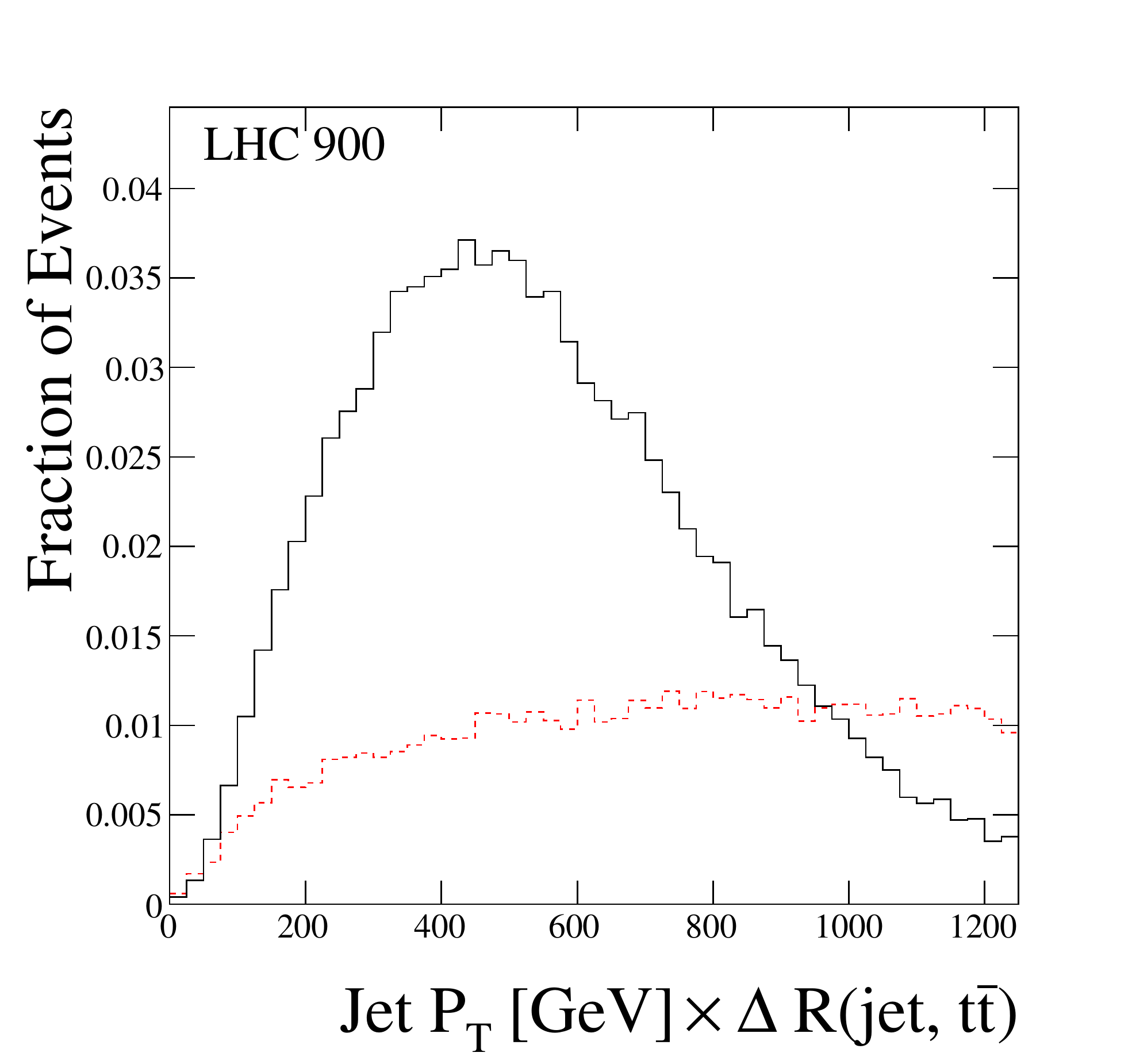}
\end{center}
\caption{ Distribution of $\Delta R(jet,t\bar{t}) \times P_T^h$ for jets from the
  initial state (red,dashed) compared to the additional jet in
  $Z'\rightarrow t\bar{t}j$ decays (black, solid). Top row is
  Tevatron; bottom row is LHC. Left column is $m_{Z'}=500$ GeV; right
  column is $m_{Z'}=900$ GeV.}
\label{fig:dr}
\end{figure}

The mass of the candidate $Z'\rightarrow t\bar{t}j$ is reconstructed
by selecting an additional jet not included in the $t\bar{t}$
assignment. With the exception of high-mass ($m_{Z'}>700$ GeV) cases at the Tevatron, the additional jet
from $Z'$ decay tends to have smaller transverse momentum than the
associated jet in $Z'+g$ or $Z'+q$ production. In addition, with the
same exception, the additional jet from $Z'$ decay tends to be close
to the $t\bar{t}$ system in angular space, see
Figure~\ref{fig:dr}. This is due to eq.~(\ref{eq:1}), which gives an enhanced coupling to
highly virtual gluons and corresponding large invariant mass of the
$t\bar t$ system, leaving the remaining jet with a relatively small
momentum. In the same way, the associated jet in the $Z'+j$ production
is preferentially at large invariant mass with the $Z'$, if allowed by
the parton luminocities.

\begin{figure}[floatfix]
\begin{center}
\includegraphics[width=0.45\linewidth]{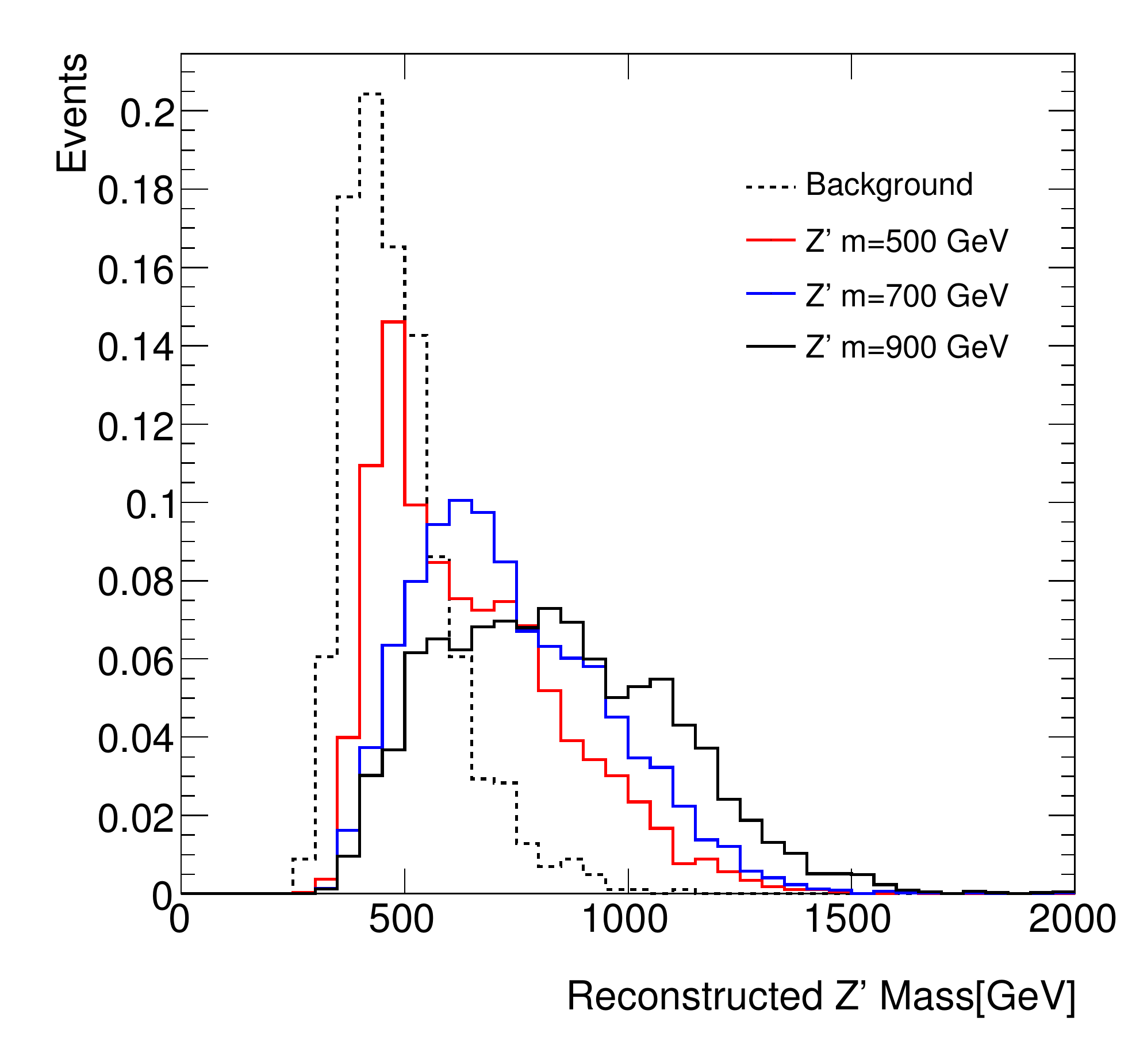}
\includegraphics[width=0.45\linewidth]{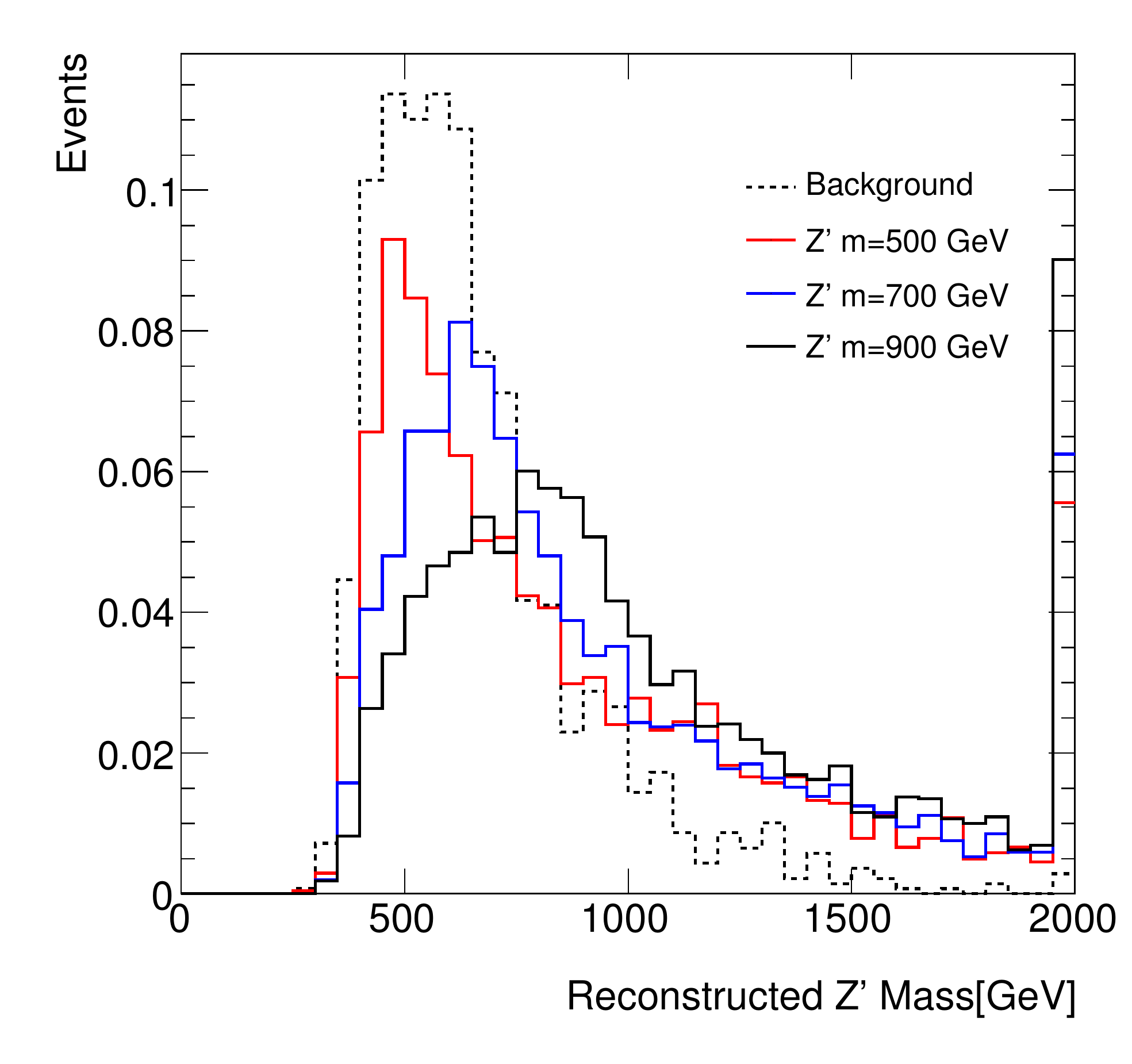}\\
\end{center}
\caption{ Distribution of reconstructed $Z'$ candidate mass
  ($m_{t\bar{t}j}$) for the dominant SM background
of $t\bar{t}$+jets and  for three choices of $Z'$ signal, using the
`near jet' as defined in the text. Overflow
  events are included in the last bin.
Left is Tevatron, right is LHC.}
\label{fig:mzp_min}
\end{figure}

\begin{figure}[floatfix]
\begin{center}
\includegraphics[width=0.45\linewidth]{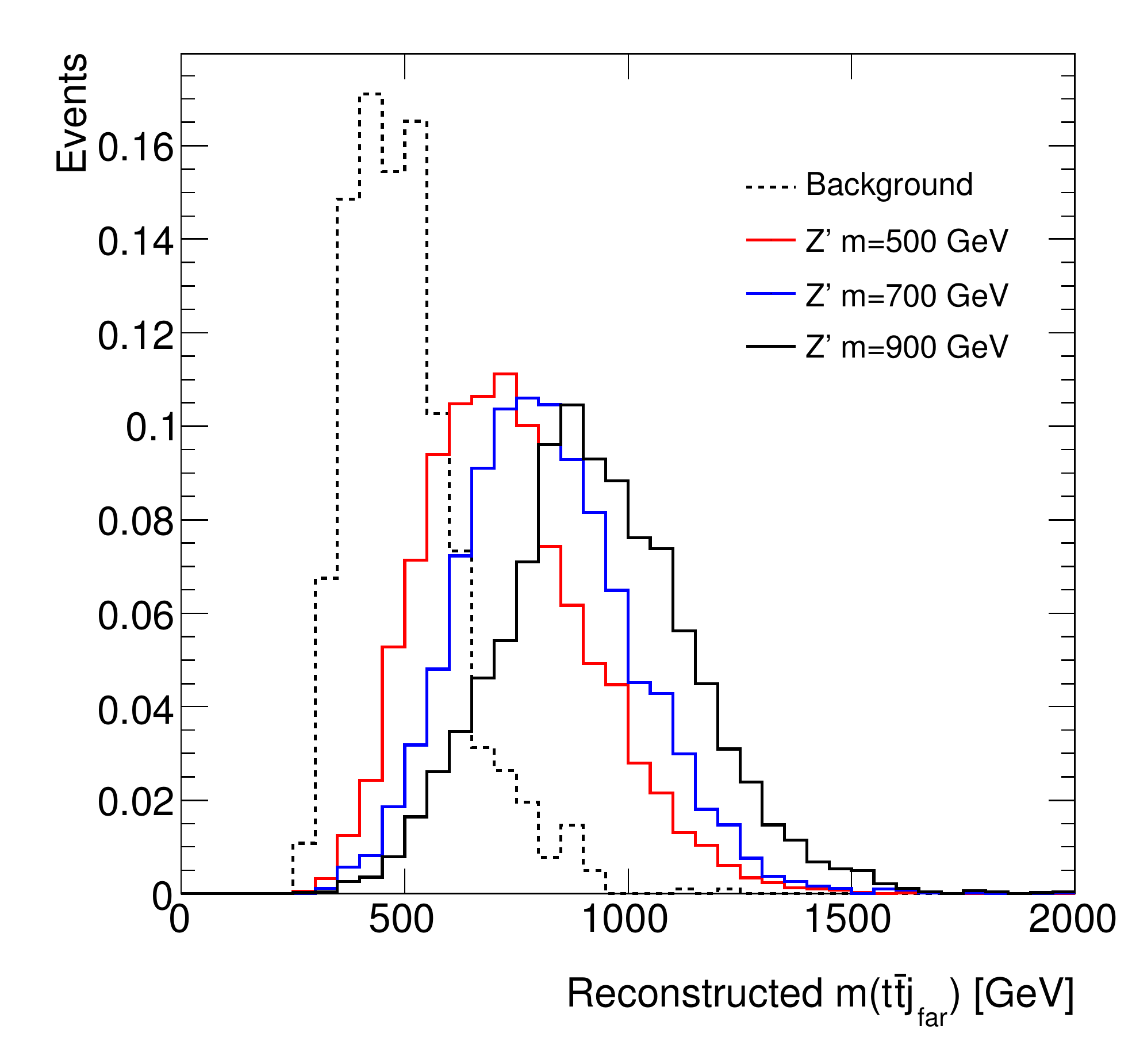}
\includegraphics[width=0.45\linewidth]{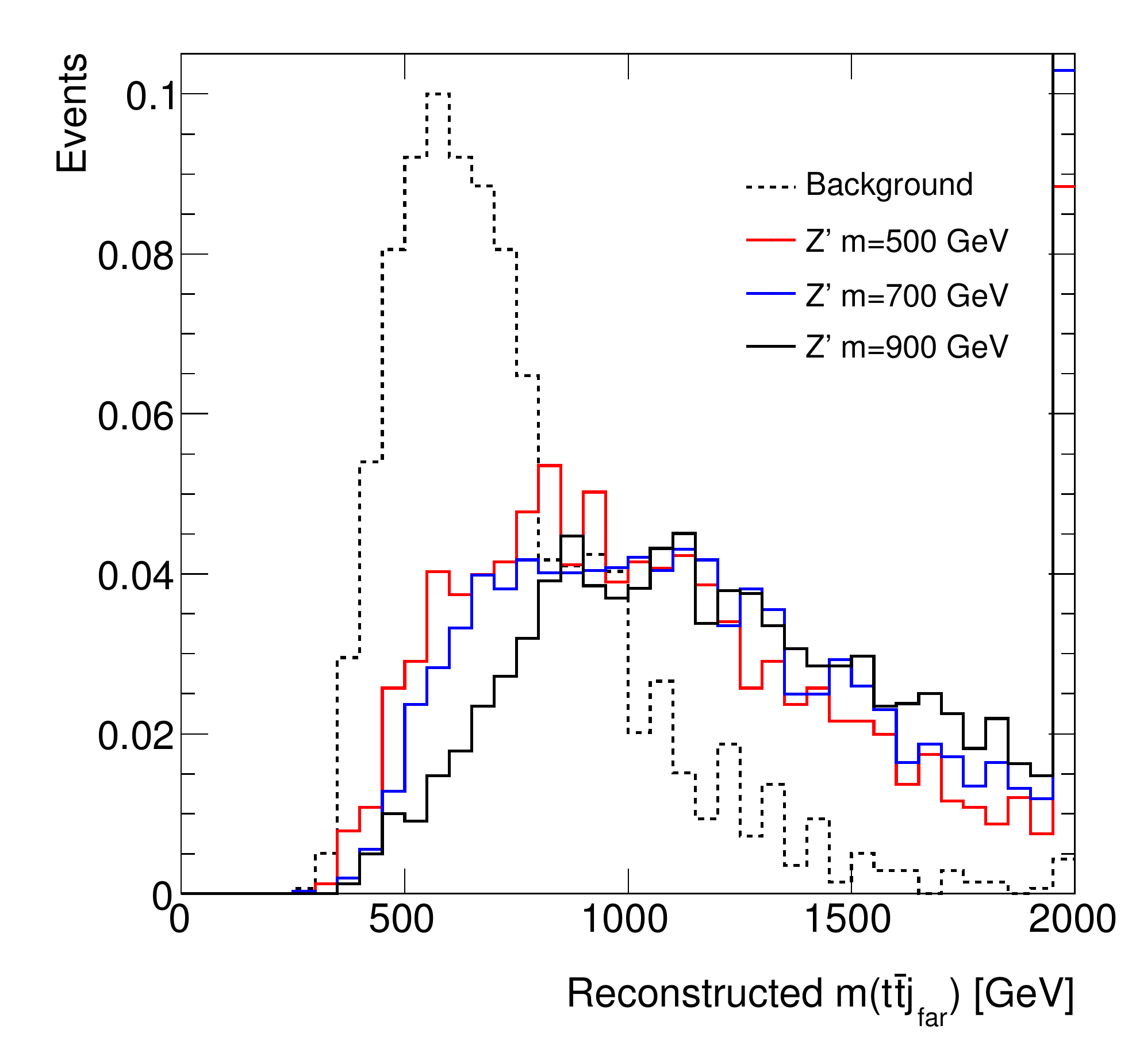}\\
\end{center}
\caption{  Distribution of the mass of the $t\bar{t}j_\mathrm{far}$ system 
  ($m_{t\bar{t}j}$) for the dominant SM background
of $t\bar{t}$+jets and
  for three choices of $Z'$ signal, using the
`far jet' as defined in the text. Overflow
  events are included in the last bin. Left is Tevatron, right is LHC.}
\label{fig:mzp_max}
\end{figure}

We therefore reconstruct the $Z'$ mass, $m_{t\bar{t}j}$
using the jet with the smallest value of  $\Delta R(j,t\bar{t}) \times
P_T^{j}$ (the `near jet'), as well as the combination $t\bar t j_\mathrm{far}$
with the jet with the largest value of  $\Delta R(j,t\bar{t}) \times
P_T^{j}$ (the `far jet')
as shown in Figures~\ref{fig:mzp_min} and ~\ref{fig:mzp_max}).  As expected, with the exception of the
high mass ($m_{Z'}>700$ GeV) case at the Tevatron, the near jet gives the most faithful reconstruction of the $Z'$ mass,
while the far jet  gives the best signal-background discrimination.

\begin{figure}[floatfix]
\begin{center}
\includegraphics[width=2.5in]{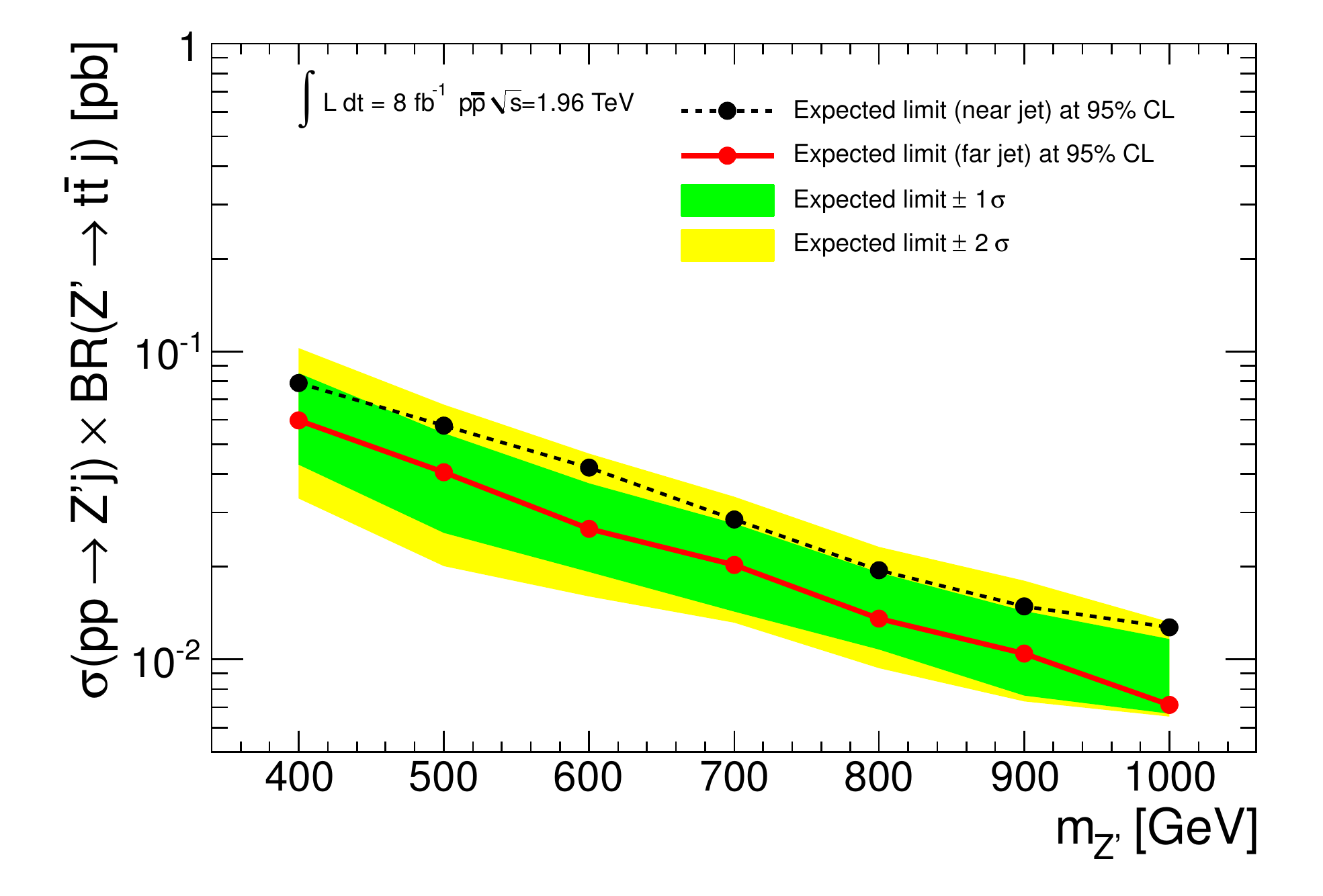}
\includegraphics[width=2.5in]{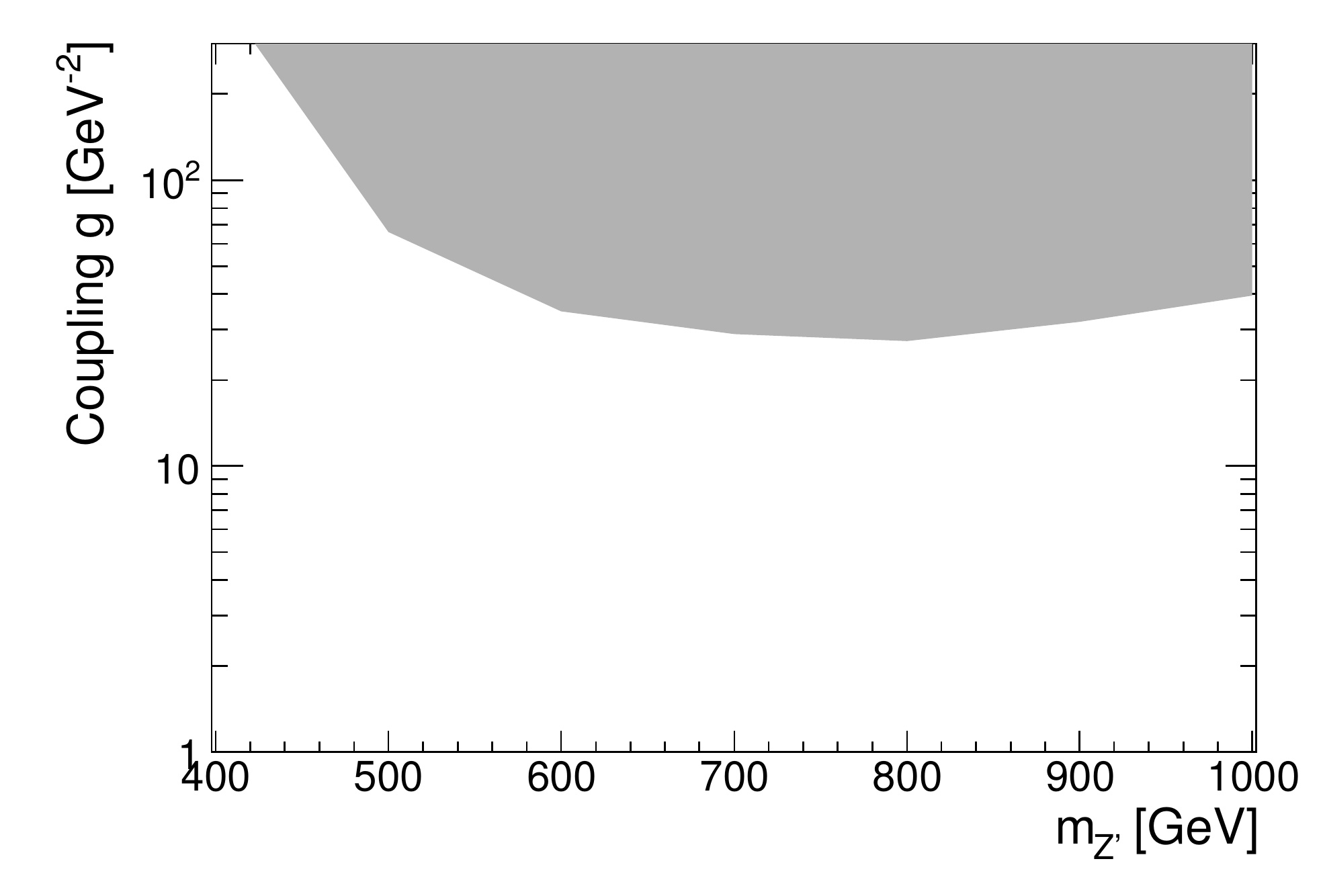}
\end{center}
\caption{ For Tevatron: top, expected upper limits on the production of $Z'+j$ at 95\%
  confidence level, as a function of $Z'$ mass; bottom, expected
  exclusion region in $m_{Z'}$ and the coupling, assuming the cross
  section has a quadratic dependence on the coupling.}
\label{fig:sens_tev}
\end{figure}

\begin{figure}[floatfix]
\begin{center}
\includegraphics[width=2.5in]{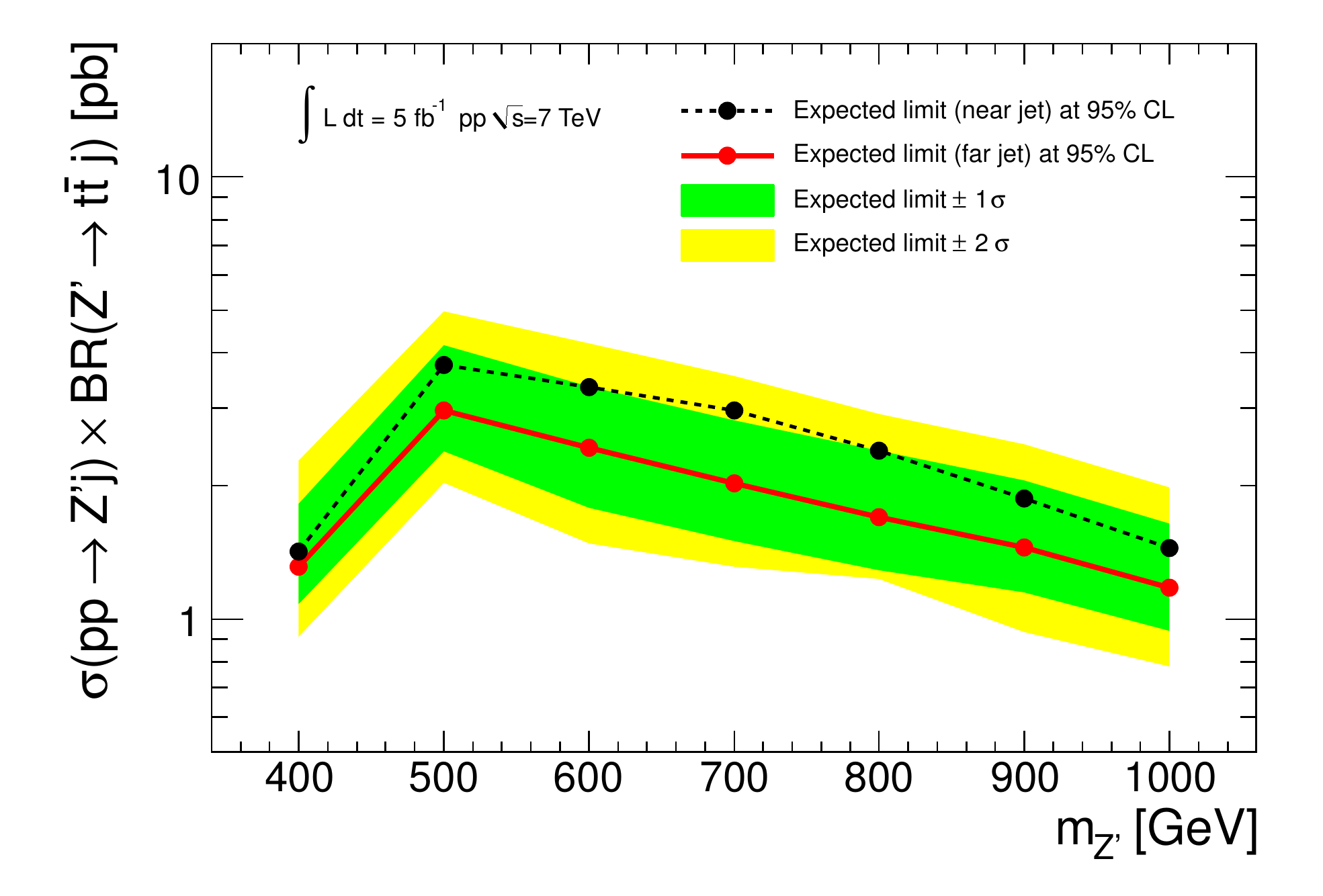}
\includegraphics[width=2.5in]{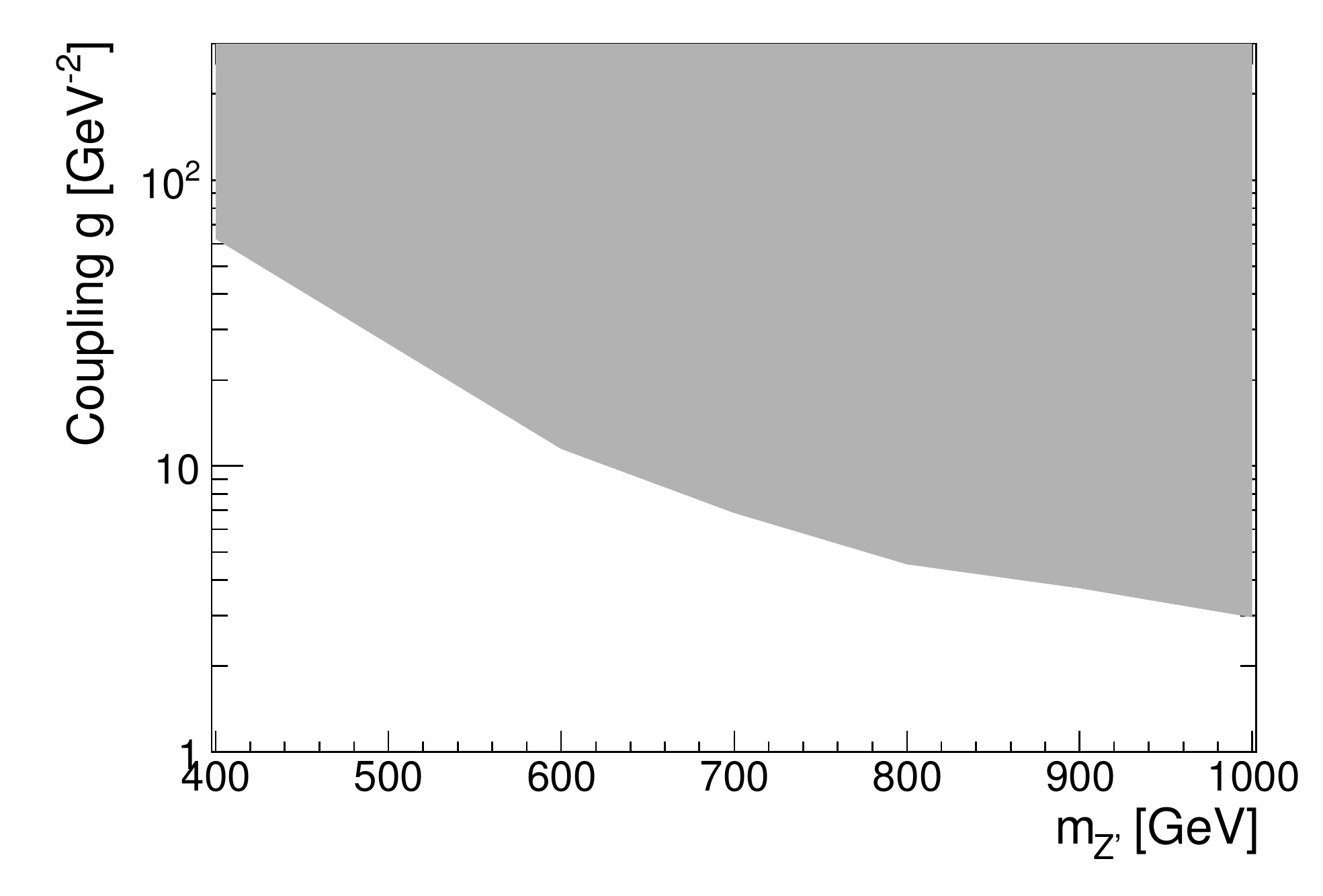}
\end{center}
\caption{ For LHC: top, expected upper limits on the production of $Z'+j$ at 95\%
  confidence level, as a function of $Z'$ mass; bottom, expected
  exclusion region in $m_{Z'}$ and the coupling, assuming the cross
  section has a quadratic dependence on the coupling.}
\label{fig:sens_lhc}
\end{figure}

To extract the most likely value of the signal cross section, a binned
maximum likelihood fit is used in the $m_{t\bar{t}j}$ variable, floating the background rate within
uncertainties. Both near- and far-jet masses are considered.
The signal and background rates are
fit simultaneously. The $\mathrm{CL_s}$ method~\cite{cls} is used to set 95\%
cross-section upper limits.  The median expected upper limit is
extracted in the background-only hypothesis, see
Figures~\ref{fig:sens_tev} and ~\ref{fig:sens_lhc}.  The far-jet mass
gives superior expected exclusion limits.

The Tevatron dataset can exclude $Z'+j$ production at the level of
$10-100$ fb in the mass range of $m_{Z'} = 400-1000$ TeV. The LHC
limits are expected to be weaker, due to the larger SM $t\bar{t}$
backgrounds. However, the expected cross-section is also larger at the
LHC. This becomes clear when the limits are expressed in the plane of
$m_{Z'}$ vs coupling $g$, assuming $\sigma(g) \propto g^2$.

\section{Conclusions}

We have introduced a model for a new heavy vector boson $Z'$, which
couples only to gluons, but may only decay via three-body decays as
$Z'\rightarrow q\bar{q}g$. This model has the additional peculiar
feature of a hard associated jet from the initial state. In the case
of top-quark decays, the signature is a resonance in $t\bar{t}+j$ with
an associated hard jet, which has not yet been experimentally
explored and to which current experimental datasets have sensitivity. We proposed two reconstruction techniques, one using a `far'
jet to establish the presence of a signal and one using a
`near' jet to perform mass reconstruction in the case of an excess. 

\section{Acknowledgements}

We thank J.~Kumar and D.~Yaylali for useful
conversations.  DW, MK and MY are supported by grants from the
Department of Energy Office of Science and by the Alfred P. Sloan Foundation.
AR is supported in part by NSF Grant No.~PHY--0653656. JA is supported
by NTU Grant number 10R1004022.

\end{document}